\documentclass[10pt,preprint]{aastex}

\journalid{}{2007} \slugcomment{To be submitted to ApJ Letter}

\newcommand{\msun}{M$_\odot$ }
\newcommand{\lsun}{L$_\odot$ }
\newcommand{\degree}{$^\circ$}
\newcommand{\degrees}{$^\circ$ }
\newcommand{\arcsecs}{$^{\prime\prime}$ }

\begin{document}

\title{Disks around massive young stellar objects: are they common? }

\author
{Zhibo Jiang\altaffilmark{1,6}, Motohide Tamura\altaffilmark{2,5}, Melvin G. Hoare\altaffilmark{3},
Yongqiang Yao\altaffilmark{4},\\ Miki Ishii\altaffilmark{5}, Min Fang\altaffilmark{1}, Ji
Yang\altaffilmark{1} } \altaffiltext{1}{Purple Mountain Observatory, 2 West Beijign Road, Nanjing
210008, China}
\altaffiltext{2}{National Astronomical Observatories of Japan, Osawa 2-21-1, Mitaka,
Tokyo 181-8588, Japan}
\altaffiltext{3}{School of Physics \& Astronomy, University of Leeds, LS2
9JT, UK}
\altaffiltext{4}{National Astronomical Observatories of China, 20 A Datun Road, Beijing
100012, China}
\altaffiltext{5}{Subaru Telescope, 650 N. A$^\prime$ohoku Place, Hilo, HI 96720,
USA}
\altaffiltext{6}{To whom reprints should be requested; E-mail: zbjiang@pmo.ac.cn}

\shortauthors{Jiang et al.} \shorttitle{Disks are common}

\begin{abstract}

We present K-band polarimetric images of several massive young
stellar objects at resolutions $\sim$ 0.1-0.5 arcsec. The
polarization vectors around these sources are nearly
centro-symmetric, indicating they are dominating the illumination of
each field. Three out of the four sources show elongated
low-polarization structures passing through the centers, suggesting
the presence of polarization disks. These structures and their
surrounding reflection nebulae make up bipolar outflow/disk systems,
supporting the collapse/accretion scenario as their low-mass
siblings. In particular, S140 IRS1 show well defined outflow cavity
walls and a polarization disk which matches the direction of
previously observed equatorial disk wind, thus confirming the
polarization disk is actually the circumstellar disk. To date, a
dozen massive protostellar objects show evidence for the existence
of disks; our work add additional samples around MYSOs equivalent to
early B-type stars.

\end{abstract}


\section{Introduction}
With the wide availability of sub-arcsec observations, the presence
of disks around massive young stellar objects (MYSOs) has become a
hot topic recently. The reason for this, is the dispute about the
way how massive stars are formed. That massive stars are formed in a
scaled-up version of how low-mass stars form \citep{shu87} would be
a natural thought. However, it has been proposed that, if the mass
of the central star exceeds 8 \msun, the tremendous radiation
pressure from the central star will halt the mass accretion and keep
it from growing \citep[e.g.,][]{kahn74,pal93}. Such a consideration
leads to an entirely different mechanism, i.e., massive stars may be
formed through mergers of lower-mass protostars \citep{bon98}. On
the other hand, a number of solutions have been proposed to account
for the radiation pressure problem, such as non-isotropic
accretion\citep{york02}, different dust opacity \citep{kahn74}, and
redirection of radiation by disk/outflow system\citep{kru05}. The
presence of disk/outflow system around MYSOs will provide key
evidence to evaluate these two kind of scenarios.

Disks around MYSOs have been detected over a wide range of
wavelengths with different tracers
\citep{pat05,jiang05,ces06,bel06}. These results suggest that up to
early B type stars \textbf{can} be formed through a similar route as
solar type stars. In this work we report sub-arcsec near-infrared
(NIR) polarimetric imaging observations which show three
polarization disks that can be best interpreted as disks or toriods
around the MYSOs.

\section{Observations and data reduction}
S140 IRS1 was observed on 2005 June 28, and the other three were
observed on 2006 November 5, all with the Coronagraphic Imager with
Adaptive Optics \citep[CIAO,][]{tam00} mounted on the Subaru
telescope. The polarization mode was set on \citep{tam03} by placing
a rotatable half-wave plate (images taken at waveplate angles of 0,
22.5, 45 and 67.5 degrees), upstream of the adaptive optics (AO)
system, and a cooled wiregrid polarizer inside the CIAO cryostat.
For each waveplate angle, the exposure times were 5 min for S140
IRS1, 4 min for S255 IRS1, 2 min for NGC7538 IRS1 and 6 min for
IRAS23033+5951, respectively. The nights were clear and seeing was
stable so that the point spread function of the optical system did
not change much. For S140 IRS1, AO was not used due to a technical
problem, resulting in a natural seeing, i.e., $\sim$ 0.45\arcsecs
(FWHM); for the other targets, the AO system was applied, delivering
a resolution of 0.15\arcsecs to 0.2\arcsec. The pixel size was set
to 0.021\arcsecs for all the observations.

The images were reduced in the standard way with IRAF packages, as
described in  \citet{jiang05}. The polarization data of S140 IRS1
were calibrated with AFGL 2591 \citep{tgj91}; whilst the others were
calibrated with NGC 7538 IRS1 \citep{yao99}. Two schemes were
applied to calculate the zero-angle correction of polarization. One
is to measure the polarization of the point sources of calibration,
with the same aperture as done in \citet{tgj91} and \citet{yao99}.
The other is to calculate the centroid of the polarization pattern
of the assumed reflection areas. By adjusting the zero-angle
correction, we can align the centroid to the center of the assumed
illuminating source. In both schemes, the zero-angle correction
agreed within one degree. Meanwhile, the assumption of
centro-symmetric polarization pattern around NGC 7538 IRS1 can be
tested by checking the non-calibrated polarization angles, which are
roughly equal in every radiating line from the object. The
polarization degrees were not calibrated because we used relative
flux to calculate them. The instrumental polarization is negligible
(less then 1\%), and the polarization efficiency at K-band is about
97\%.

\section{Results}
The main results are shown in Fig. 1. A common feature is noticed
from inspection of the polarization vectors (left panels) for each
field. In most of the fields, the polarization vectors are circling
around the center, forming centro-symmetric patterns, suggesting the
MYSOs are dominating the illumination of some or all of the fields.
However, details of the polarization morphology are different from
region to region, and are described below.

\textbf{S140 IRS1}: The polarization vectors are perfectly
centro-symmetric (Fig. 1a, left panel), suggesting that the source
is dominating its surrounding area; no other sources affect the
polarization. Two low-polarization lanes pass through the central
source. One is in the southwest-northeast at a position angle (p.a.,
measured east of north) $\sim$ 45\degree, which is rather straight
and narrow; the other is wider and a little curved to the north, and
virtually perpendicular to the first one. In the pseudocolor image
composed from intensity image and polarized flux image(Fig. 1a,
right panel), two parabolically curved features, separated by the
southwest-northeast lane, presumably representing reflection
nebulosities, are seen opening towards the southeast and northwest,
respectively. These two reflection nebulosities is probably the
outflow cavity walls. A scattered photon is effectively polarized if
the scattering angle (the angle between the incident and emerging
direction) is near 90\degrees \citep[e.g.,][]{cod95}. This is why we
can only see the polarized light on two sides of the wall, while
little polarization is observed in between. It is particularly true
for large opening angles, as in the case of S140 IRS1. The wall
profile to the northwest is more clearly observable, because it
delineates a red-shifted outflow lobe \citep{hay87,min93}, where
extinction is larger so that we can only see the most efficiently
polarized light on the two sides. The blue-shifted outflow to the
southeast shows much brighter reflection nebulosity, consistent with
the result from the NIR speckle-interferogram images \citep{sch00,
wei02}. The asymmetry of the brightness between the nebulae, which
had also been observed in CO lines \citep{hay87}, is probably a
result of dust distribution around this object. The morphologies of
the walls indicate that the opening angle is rather large
($\sim$150\degree), consistent with the poor outflow collimation
ratio \citep{hay87}.

The low-polarization lane in the southwest-northeast, therefore,
represents a polarization disk (PD). Such a kind of feature is
observed in intermediate-mass young stars \citep{per04} as well as
in low- and high-mass protostars \citep{luc04,jiang05}.
\citet{mel06} observed IRS1 in the centimeter continuum band, and
found an elongated structure, which is interpreted as an equatorial
disk wind. An over-plot of the ionized disk wind on the pseudocolor
image shows that the directions of the PD and the disk wind matches
each other (Fig. 1a, right panel). This evidence strongly suggests
that the PD is actually a physical disk. The diameter of the PD is
rather large ($\ge$2700 AU), implying the dusty disk is more
extended than the ionized disk wind. Based on the bow-shock-like
structures in the K$^\prime$-band image, \citet{wei02} proposed a
quadrupolar outflow scenario for IRS1 \citep[see also][]{yao98}.
However, since VLA4 is located along the path that the bow-shocks
trace \citep[Fig. 2,][]{wei02}, it is also possible that the
20\degrees outflow is driven by VLA4. Given the configuration of the
system we observed, a single bipolar outflow/disk system for IRS1 is
more likely than multiple outflows.

\textbf{S255 IRS1}: A bipolar nebula is located around S255 IRS1
(Fig. 1b, right panel) at p.a. $\sim$35\degree. High degrees of
polarization (up to $\sim$30\%) in the nebula indicate that it
should be a reflection nebula illuminated by the central source,
IRS1, as illustrated by the polarization vectors (Fig. 1b, left
panel). In the southwest part of the nebula there could be some
contaminations--the outflow from IRS3 may overlap \citep{tgj91}.
Still farther to the south ($\sim$ 3\arcsec from IRS1), a bright
nebula is seen with scattering dominated by IRS3. In spite of this,
the morphology of the bipolar nebula suggests an outflow from IRS1
with a small opening angle ($\le 50^\circ$). Additional evidence is
the H$_2$ emission knots \citep[S255:H2 2,][]{mir97}, which is
located along the extension of the southern nebula (p.a. $\sim$ 220
\degree), and which has been suggested to be excited by IRS1. The
nebula is a little twisted in ``S'' shape, suggesting that it is a
precessing outflow.

A biconical low-polarization lane runs through IRS1, roughly
perpendicular to the axis of the reflection nebula at p.a.
$\sim$110\degree, with a length of $\sim$5700 AU. In Fig 1b (right
panel) we indicate the shape of the structure between two curves.
The polarization vectors there are roughly parallel to the lane, but
the whole pattern of vectors is approximately elliptical. This
structure is typically a PD, which has been well modelled
\citep[e.g.,][]{bas88} and observed in many cases
\citep[e.g.,][]{whi97}. It usually suggests a disk or a toriod
around the central source.

\textbf{NGC7538 IRS1}: Bright nebulosities are seen to the north and
northeast of IRS1, but rather faint to the SE (Fig. 1c, right
panel). The polarization degrees to the north (up to 30\%) are
higher than to the south ($\sim 10\%$), and the polarized flux there
show irregular structures. The polarization pattern of these
nebulosities is generally centro-symmetric with respect to IRS1
(Fig. 1c, left panel). This confirms the suggestion by \citet{kra06}
that the bright patches to the northwest are illuminated by IRS1.
Within $\sim$1\arcsecs of IRS1, the polarization degrees are lower
than the outside area, with a few low-polarization patches radiating
out. There could be a PD surrounding this source, but we cannot tell
which one it is from the complicated polarization morphology.

From radio observations \citet{gau95} suggest a south-north outflow from IRS1. The reflection
nebulosities to the northwest and north are therefore likely to be the cavity wall of the
blue-shifted outflow. Previous observations \citep{cam84,dav98} show the redshifted lobe is
spatially coincident with the blue-shifted one which extends to the northwest, suggesting that the
outflow is seen nearly pole-on. This would explain why we do not clearly see the southern
counterpart of the reflection nebulosity where the extinction is much larger.

From the methanol maser observation, \citet{pes04} suggested that there is a disk in the
southeast-northwest direction. However, methanol masers can also be excited by high velocity gas.
In light of our observation, it is also possible that the feature observed by Pestalozzi et al.
represents one side of the cavity walls. Such an interpretation does not conflict to the scenario
of a precessing outflow \citep{kra06}.

\textbf{IRAS 23033+5951}: This is the most complicated region of the
four. In the pseudocolor image (Fig. 1d, right panel), a conical
nebulosity is seen, opening to the east. Several peaks are detected,
but only the one marked by a plus (S1) is point-like (FWHM $\sim
0.3\arcsec$). To the east of S1, two other major peaks are detected.
That they are not point-like and they have high polarizations
suggests they are infrared reflection nebulosities (IRN).  The
northern one (IRN1) is mainly reflective; the polarization degrees
are high ($\sim$20 \%) and the vectors are facing towards S1,
indicating it is illuminated by S1. The polarization degrees are
lower ($\sim$10 \%) in the southern one (IRN2), but the vectors also
face to S1. The lower polarization could be caused either by a
mixture of reflected light and self-radiation or by multiple
illumination.   To the northwest of S1, a faint reflection
nebulosity is noticeable (IRN3), with relatively high polarization
degrees ($\sim$30 \%). The polarization vectors suggest this
nebulosity is illuminated by another probably deeply embedded
protostar that is not visible in the NIR. Since the property of this
source is beyond the scope of this paper, a detailed analysis of it
will be presented later (Fang et al., in preparation).

Again, we see a low-polarization lane, which is more clearly seen in
the polarization degree image (Fig 1d, left panel), running roughly
south-north (p.a. $\sim$160\degree) and passing through the center
of S1; the polarization vectors are generally aligned with the
direction of the lane. As discussed above, this lane suggests the
presence of a disk or a toroid. The length of this structure is
about 6700 AU. The reflection nebulosities (IRN1 and IRN2) that are
illuminated by S1, thus represent one direction of the outflow
driven by S1. This result is quite intriguing since a bipolar
outflow has been detected in the southeast-northwest \citep{kum02}.
Probably this outflow is driven by the deeply embedded protostar
mentioned above. On the other hand, the outflow driven by S1 has its
own manifestation. A shocked H$_2$ emission knot is found 4\arcsecs
west of S1 \citep{kum02}, which is presumably excited by S1. It is
interesting to note that the illuminating source, S1, is fainter
than its illuminated reflection nebulosities in the K-band. A
possible explanation is that the disk is seen edge-on so that the
light directly from the central source is heavily extincted while it
can easily escape from the outflow cavity, then being reflected by
the cavity walls. We note however, if only the extinction toward
IRN1 and IRN2 is larger than that toward S1
(A$_{V(IRN1)}$-A$_{V(S1)}$ $\geq$15), an observed effect can be
reproduced.

\section{Implications and conclusions}

There are several possible causes that make a low-polarization lane
passing through the bright source. In addition to a flattened
structure around the illuminating source, a magnetic field may also
produce a similar morphology by dichroic extinction. We checked the
polarization of point sources in our observed fields, and find no
large-scale alignment of polarization vectors in the direction of
PDs. Another possibility is that, when a nearly spherical envelop
illuminated by a source inside and a much brighter source outside,
the envelop will show a bipolar polarization structure. However,
this happens when the source outside is hundreds to thousands of
times brighter than the source inside, so that the photons from the
two sources shining on the envelop are roughly equal. It is not
possible in our case since our sources are brighter than 10$^3$
\lsun. It is also possible that a foreground filament which is just
in front of the sources dilutes the polarization, but it is of very
low probability and is in fact impossible to happen for all three
targets. Therefore, taking the morphologies of the observed
reflection nebulae into consideration, the most likely
interpretation of the PDs here would be real disks or toriods. We
note however, since the polarimetric observations do not give
velocity information, it is not possible to distinguish between a
Keplerian disk and a toriod structure. A confirmation of the
conclusion have to await sensitive molecular line observations.

Some parameters of the sources are listed in Table 1.  Of the four
sources, three (S140 IRS1, S255 IRS1 and I23033) show evidence for
the presence of a disk or toroid.  All of them are deeply embedded
(A$_V$ $>$ 15 for IRAS23033+5951, $>$20 for S255 IRS1 and $>$30 for
S140 and NGC7538, Fig. 2a). They also show very large near infrared
color excesses. Their near infrared colors suggest the evolutionary
status could be comparable to low-mass Class I objects. Except for
S255 IRS1, whose spectral type is still controversial
\citep{hey89,how97,ito01}, these sources are MYSOs with masses
ranging between 10 to 30 solar masses according to their spectral
types. A J vs J-H plot would also suggest a similar conclusion about
spectral types of these objects (Fig. 2b).

Up to date, a dozen of MYSOs show evidence of the existence of disks
\citep{ces06,zin07}. These candidate disks are found in various
evolutionary status of their host objects, which are detected from
(sub)millimeter wavelengths the NIR. However, the candidate disks
happen around MYSOs with their luminosities less that 10$^5$ \lsun,
typical of early B-type main-sequence stars \citep{ces06},
suggesting that disks might be common occurrence for this kind of
MYSOs; we still lack of evidence whether still higher-mass YSOs host
circumstellar disks. Our observations add samples to this kind. And
it is interesting to note that our most massive sample, NGC7538 IRS1
($\sim$ 30 \msun), does not show strong evidence of PD.

We have also shown that high-resolution polarimetric imaging can
provide a sensitive tool to detect circumstellar disks around MYSOs,
even if the disks are too faint to be detectable directly. Our
future work will concentrate on searching for polarization disks
around MYSOs of higher masses and different evolutionary status. We
aim to set up a sample for future observations with higher
resolution and sensitivity. This kind of work will give crucial
hints to answer the question at what mass and evolutionary phase the
disks around MYSOs are truncated.

\acknowledgments

This work is based on data collected at Subaru Telescope, which is
operated by the National Astronomical Observatory of Japan. We
acknowledge the staff of the Subaru telescope for their helps during
the observations. This work makes use of the SIMBAD database,
operated at CDS, Strasbourg, France, and the 2MASS data base, which
is funded by the National Aeronautics and Space Administration and
the National Science Foundation. This work is supported by a
Grant-in-Aid from MEXT Nos 16077204 and 16077171, Japan, and NSFC
Nos 10473022 and 10621303, China, and partially supported by
Ministry of Science and Technology (2007CB815406) of China.

\begin{deluxetable}{cccccc}
\tablecolumns{8} \tablewidth{0pc} \tabletypesize{\scriptsize}
\tablecaption{Parameters of the sources} \tablehead{ \colhead{Name}
&   \colhead{D(kpc) } &  \colhead{L(\lsun)} &    \colhead{Sp type }
&   \colhead{Size(AU)\tablenotemark{1}} &
  \colhead{ Ref.\tablenotemark{2}}}
\startdata
 S140 IRS1       &       0.9  &   5.00E+03 &    early B & 2700 &  (1)    \\
S255 IRS1        &       2.4  &   \nodata    &       O,B0 & 5700 & (2,3)    \\
NGC 7538 IRS1    &       2.8  &   8.30E+04 &         O6 & \nodata  &  (4)    \\
IRAS 23033+5951  &       3.5  &   2.51E+04 &       B0.5 & 6700 & (5)     \\
\enddata
\tablenotetext{1}{The projected length of the PD at adopted distance.}  \\
\tablenotetext{2}{Ref. Code 1. \cite{les86}; 2. \cite{how97}; 3. \cite{ito01};  4. \cite{kra06}; 5.
\cite{wil05}}
\end{deluxetable}

\clearpage

\normalsize
\begin{figure}
\includegraphics[totalheight=6.5in,angle=0]{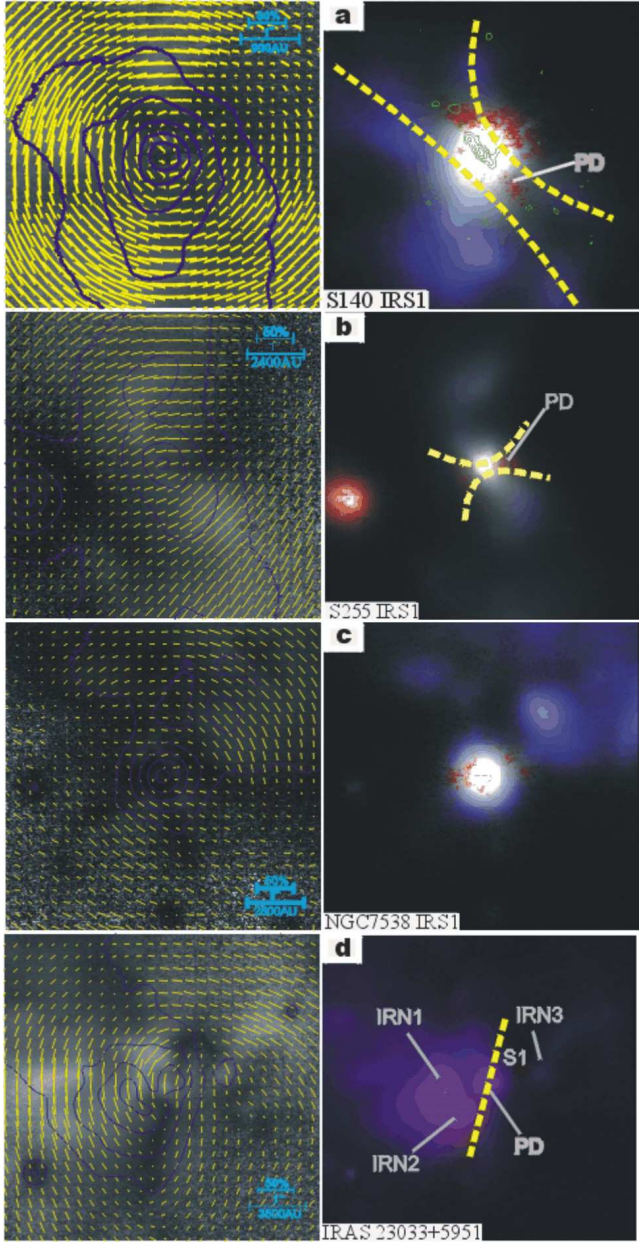}
\caption {(left panels) Polarization degree images overlaid by
polarization vectors (yellow dashes) and total intensity contour
(blue curves). The angular and polarization scales are shown in the
corners. The contours start from 10.93, 14.56, 11.34, 12.42 mag
arcsec$^{-2}$, respectively, and decrease every 1.25 mag
arcsec$^{-2}$. (right panels) Pseudocolor images composed of pure
brightness images (red) and polarized brightness images (blue). A
blue feature indicates the highly polarized nebulosity, which should
be reflective. Red features are low polarization areas. The PD is
indicated with double lines to delineate the edge of the disks (S140
IRS1 and S255 IRS2), or a single line to indicate the position
(IRAS23033+5951). For S140 IRS1, The contours of 5 GHz continuum
emission representing the ionized equatorial disk wind are
superimposed \citep{mel06}. North is up and east is to the left.}
\end{figure}

\begin{figure}
\includegraphics[totalheight=5.0cm,angle=0]{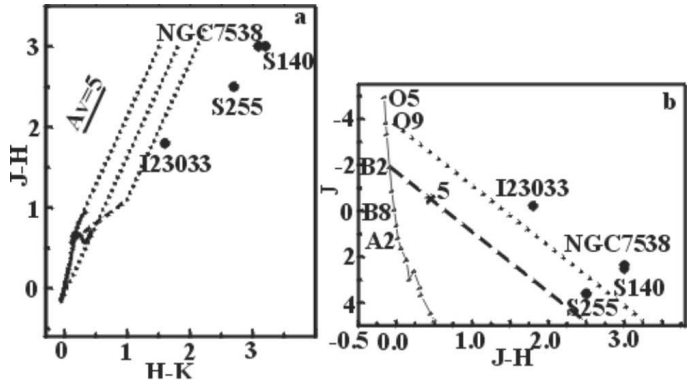}
\caption {a. Near infrared color-color diagram. The JHK magnitudes
were obtained from the 2MASS database. Solid curves show the loci of
main-sequence dwarfs and giants. The dotted lines indicate the
reddening bands of main-sequence and T-Tauri stars.  b. J vs J-H
color-magnitude diagram of the sources. The solid curve indicates
the zero-age main sequence stars with different spectral type. The
dashed and dotted lines are drawn from B2 and O9 stars, parallel to
the extinction vector, the scale of which is indicated by a cross on
the dashed line. }
\end{figure}

\end{document}